\documentclass{article}

\usepackage{arxiv}
\usepackage[utf8]{inputenc} 
\usepackage[T1]{fontenc}    
\usepackage{hyperref}       
\usepackage{url}            
\usepackage{booktabs}       
\usepackage{amsfonts}       
\usepackage{nicefrac}       
\usepackage{microtype}      
\usepackage{graphicx}
\usepackage[numbers]{natbib}
\usepackage{doi}

\title{AIRPET: Virtual Positron Emission Tomography}

\date{}

\author{
    J. Renner \\
	Instituto de F\'isica Corpuscular (IFIC)\\
    CSIC \& Universitat de Val\`encia\\
    Calle Catedr\'atico Jos\'e Beltr\'an, 2\\
    Paterna, E-46980, Spain\\
	\texttt{jrenner@ific.uv.es}
	\And
    J.J. G\'{o}mez Cadenas\\
	Donostia International Physics Center\\
    BERC Basque Excellence Research Centre\\
    Manuel de Lardizabal 4\\
    San Sebasti\'an / Donostia, E-20018, Spain\\\\
    Ikerbasque (Basque Foundation for Science)\\
    Bilbao, E-48009, Spain\\
	\texttt{jjgomezcadenas@dipc.org}
    \And
    S.R. Soleti\\
	Donostia International Physics Center\\
    BERC Basque Excellence Research Centre\\
    Manuel de Lardizabal 4\\
    San Sebasti\'an / Donostia, E-20018, Spain\\\\
    Ikerbasque (Basque Foundation for Science)\\
    Bilbao, E-48009, Spain\\
	\texttt{roberto.soleti@dipc.org}
}

\renewcommand{\headeright}{}
\renewcommand{\undertitle}{}

\hypersetup{
pdftitle={AIRPET: Virtual Positron Emission Tomography},
pdfsubject={},
pdfauthor={},
pdfkeywords={PET, Geant4, GDML},
}

\begin{document}
\maketitle

\begin{abstract}
Positron Emission Tomography (PET) is a powerful medical imaging technique, but the design and evaluation of new PET scanner technologies present significant challenges. The process is typically divided into three major stages: 1. detector design and simulation, 2. image reconstruction, and 3. image interpretation. Each of these stages requires significant expertise, making it difficult for individuals or small teams to manage all three at once. AIRPET (AI-driven Revolution in Positron Emission Tomography) is a web-based platform designed to address this challenge by integrating all phases of PET design into a single, accessible, and AI-assisted workflow. AIRPET provides an interface to large language models (LLMs) for assisted geometry creation and an interface for basic PET image reconstruction with the potential for further expansion. Here we introduce AIRPET and outline its current functionality and proposed additions.
\end{abstract}

\textbf{\emph{Code:}} \url{https://github.com/jerenner/airpet}

\section{Introduction}
\label{sec:introduction}

Positron Emission Tomography (PET) is an important technique in modern medical diagnostics, providing functional imaging of metabolic processes within the body~\cite{Shukla2006PET,Zatcepin2023}. The development of new PET scanners is a multidisciplinary endeavor involving three principle stages:
\begin{enumerate}
    \item[1.] \textbf{Detector Design and Simulation:} This first stage involves setting up the PET detector model, including its geometry and materials. This is done with a simulation toolkit such as Geant4~\cite{geant4,allison2016}. The relevant PET physics, including electron-positron annihilation and propagation of the resulting 511 keV gamma rays through the detector geometry, must then be simulated via Monte Carlo methods for a given radiotracer distribution. The effects of detector reconstruction such as position and energy resolution must also be added.
    \item[2.] \textbf{Image Reconstruction:} The locations of the detected gamma ray interactions, must be processed using reconstruction algorithms to obtain a 3D image of the simulated distribution. This step can be carried out using a code package such as~\cite{Merlin2018, Thielemans2012,parallelproj}.
    \item[3.] \textbf{Image Interpretation:} The reconstructed image must be evaluated from a medical standpoint to interpret the results of the scan. This step is performed by a medical professional.
\end{enumerate}

Stages (1) and (2) require a specialized skillset, creating a significant barrier to entry for researchers and engineers. To address this challenge, we have developed AIRPET, an AI-assisted platform for design and simulation of PET detectors. By integrating design, simulation, and image reconstruction into a single interface, AIRPET facilitates access to PET research and development. In addition,  AIRPET aims to provide both a training ground (generation of training sets along the lines discussed above), and an AI partner, whose diagnoses can be compared with those of radiologists. 

The design of a modern PET scanner, such as a total-body PET (TBPET) system~\cite{Vandenberghe2020}, requires careful consideration of performance, cost, and clinical utility. As detailed in the development of the CRYSP TBPET scanner~\cite{Soleti2025}, choices in scintillator materials, detector geometry, and electronics all have significant effects on the final image quality and diagnostic capabilities. The CRYSP project, for instance, explores the use of cryogenic cesium iodide crystals as a cost-effective alternative to more traditional, expensive materials, demonstrating the type of design choices that require extensive simulation and evaluation.

The current PET design landscape consists of a variety of specialized software tools~\cite{delaPrieta2012} for each stage of the process, but they require significant expertise to operate and integrate. This creates a siloed environment where experts in one domain may struggle to fully appreciate the implications of their work on the others. An integrated tool that bridges these gaps would lead to a more efficient design process.

Stage (3) is performed by a medical professional. One way in which new technologies can be useful for radiologists is by providing them with an ``AI partner'' and a ``training environment.'' While in many cases training can and will occur by examining real labeled data, AI tools combined with simulations can be used to create variants and combinations of real cases, as well as new, simulated, realistic problems for diagnosis (which are verifiable, since one has the true information).

\section{AIRPET: A Unified Platform for PET Design}
\begin{figure}[!htb]
    \centering
    \setlength{\fboxsep}{0pt} 
    \setlength{\fboxrule}{1pt}
    \fbox{\includegraphics[scale=0.31]{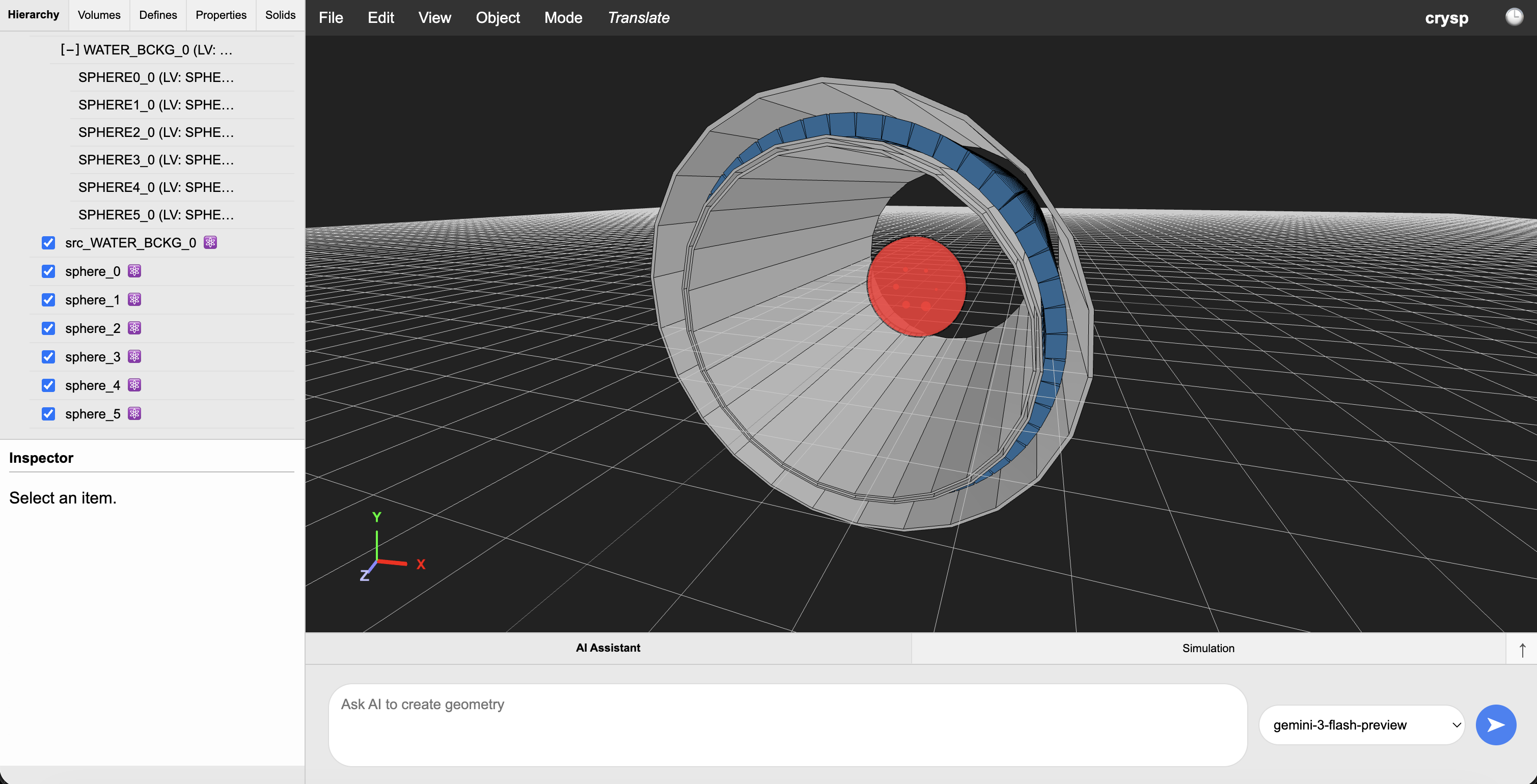}}
    \caption{Example of the geometry interface in AIRPET, showing the CRYSP1M~\cite{Soleti2025} detector geometry and visualisation of several simulated tracks.}
    \label{fig:geom}
\end{figure}

AIRPET is being developed as an open-source, web-based application that assists the user through the PET design and evaluation workflow. The project has already implemented basic functionality in detector geometry specification and allows for execution of basic simulations and PET image reconstruction. The end-to-end simulation and reconstruction workflow is as follows:
\begin{enumerate}
    \item [1.] The user defines a geometry and simulation configuration in the web-based GUI (see Fig. \ref{fig:geom}).
    \item [2.] The application generates the necessary files to launch the corresponding detector physics simulation. The resulting output can be analyzed directly by the user and can be useful even outside of PET physics, and/or the user can continue a PET physics analysis with step 3 below.
    \item [3.] The user launches the PET reconstruction interface, which allows for computation of lines of response (LORs) and execution of the 3D PET image reconstruction from those LORs. The reconstructed 3D image array is saved and can be interactively viewed slice-by-slice in the frontend (see Fig. \ref{fig:reco}).
\end{enumerate}

\subsection{Detector Design}
AIRPET operates on a client-server model in which the frontend (HTML/Javascript) serves as the user interface for geometry manipulation and simulation preparation, while the backend (Python/Flask) manages the project state, simulation, and image reconstruction.  The initial and most developed component of AIRPET is a robust toolchain focused on the creation of detector geometry. It begins with a visual editor employing three.js~\cite{threejs} for 3D rendering that allows for intuitive interaction with the geometry. It uses an internal JSON-based representation for saving and versioning of designs while also including support for import and export of the standard Geant4 GDML~\cite{chytracek2006} format. Users can add, delete, and modify components within the scope of GDML, including defines, materials, solids (both primitive and complex booleans), logical and physical volumes, and visualization attributes. 

A key feature of AIRPET is the integration of an AI assistant to allow for geometry creation via text input. Currently locally-hosted open-source models (via Ollama~\cite{ollama2023}) and cloud-based models from Google Gemini~\cite{gemini2023} are supported. A user can, for example, simply request to ``create a minimal PET geometry consisting of a ring of 16 crystals with a 90 cm radius,'' and the AI backend will attempt to translate this command into the necessary JSON structure to specify the geometry. Note that this feature in practice cannot currently replace a skilled human designer, and the quality of the responses depends heavily on the model employed and the user-specified prompt. While the results may not be perfect, even with current top-of-the-line models, AIRPET provides the tools to perform a significant fraction of the work with AI and then fine-tune the result manually. AIRPET also allows for CAD import via STEP files. This feature uses the pythonocc~\cite{pythonocc} library to parse STEP files, converting them into Geant4-compatible tessellated solids.

\subsection{Detector Simulation}
AIRPET is integrated with Geant4 via a pre-compiled simulation targeted for use with the interface. When a user initiates a run, the backend automatically generates the current geometry as a GDML file and creates a corresponding Geant4 macro (.mac) file based on the user-defined sources and simulation parameters. It then executes the simulation in a separate thread, providing real-time progress and log messages to the user's web interface. The output, including detector hits and particle tracks, is stored in HDF5 format for subsequent analysis. Users can choose to continue with the integrated PET analysis or write their own analysis based on the simulation output.

\begin{figure}[!htb]
    \centering
    \setlength{\fboxsep}{0pt} 
    \setlength{\fboxrule}{1pt}
    \fbox{\includegraphics[scale=0.35]{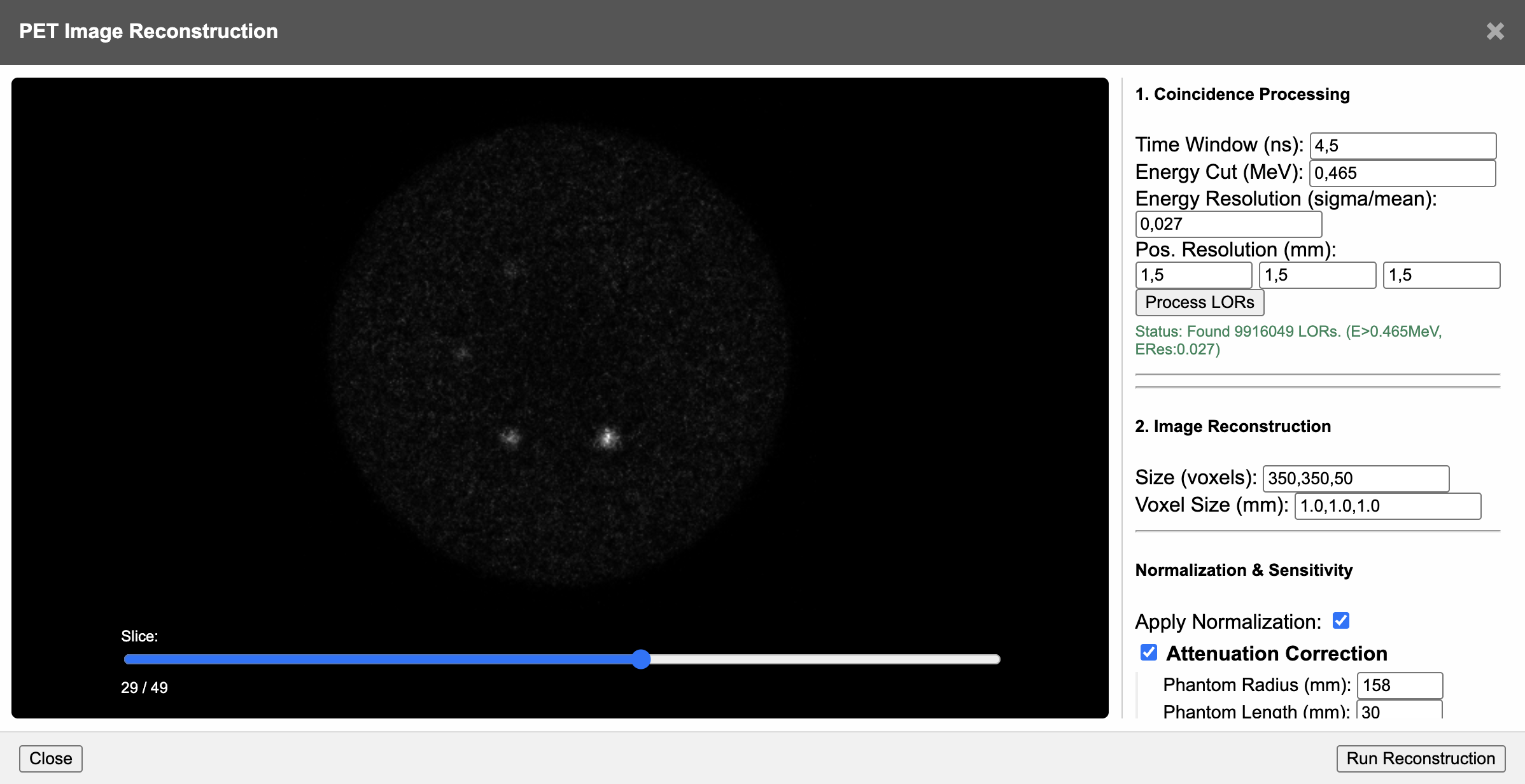}}
    \caption{Example of image reconstruction interface in AIRPET.}
    \label{fig:reco}
\end{figure}

\subsection{Image Reconstruction}
To produce a PET image, reconstructed locations of detected 511 keV gamma rays must first be processed into LORs. Detector physics effects, such as position reconstruction resolution and energy resolution, can be applied to the results of the detector simulation to produce realistic reconstructed locations. This is currently quite limited in AIRPET: initial hit locations in each sensitive volume are taken as the reconstructed locations, and a Gaussian smearing is applied to the reconstructed positions and energy. A dataset of LORs can be generated for a given simulation, which provides input for the image reconstruction.

AIRPET interfaces with the parallelproj~\cite{parallelproj} library for MLEM image reconstruction, designed to transform the raw LOR data from the simulation stage into tangible 3D PET images. This provides users with feedback on their detector's performance, closing the loop between design and imaging results. The reconstruction module currently makes use of Maximum Likelihood Expectation Maximization (MLEM) reconstruction~\cite{Shepp1982}. The reconstructed images can be rendered directly within the AIRPET interface to provide immediate visual feedback, as shown in Fig. \ref{fig:reco}.

\subsection{Medical Evaluation}
The final stage of AIRPET development will be the inclusion of a medical evaluation module employing multi-modal large language models~\cite{Fu2024} to provide a qualitative assessment of the reconstructed PET images~\cite{PintoCoelho2023}. For example, a user could present the AI with a reconstructed image of a phantom containing simulated lesions and ask for an ``opinion'' on the image's diagnostic quality.
It is important to emphasize that this feature will not be a substitute for evaluation by a qualified medical professional. The primary goal at this stage is not to provide medical advice, but rather to provide feedback on how design choices might impact the performance of a PET scanner.

\section{CRYSP: An example use case}
\noindent Here we present a brief example of image reconstruction using AIRPET, in which we model a simplified version of the CRYSP1M \cite{Soleti2025} geometry, consisting of 4.8x4.8x3.72 cm$^3$ CsI crystals arranged in rings of approx. 40 cm radius covering approx. 1 m in axial length. 

\begin{figure}[!htb]
    \centering
    \includegraphics[scale=0.55]{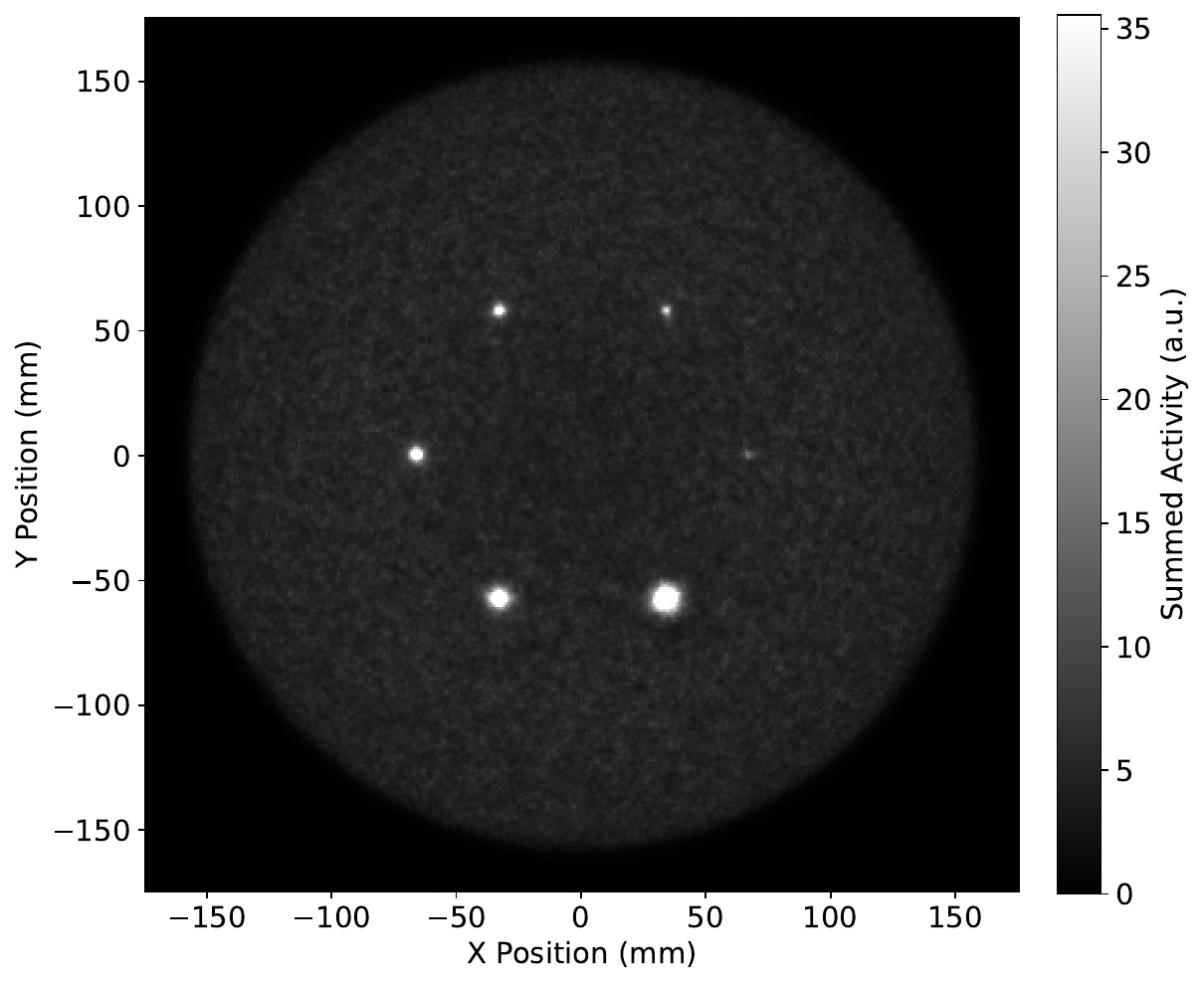}
    \caption{Reconstructed image of a 6-sphere, water-filled phantom in the CRYSP-1M geometry.}
    \label{fig:reco_img}
\end{figure}

A 6-sphere water-filled phantom was placed in the center of the device, including spheres of radii 4.75 mm, 6.35 mm, 7.95 mm, 9.55 mm, 12.7 mm, and 15.9 mm. The spheres were surrounded by a cylindrical volume of water, and a ratio of activity per unit volume of sphere/water = 4/1 was assigned. 10$^8$ e$^+$/e$^-$ decays were simulated in the active volume, and the image was reconstructed using 10 iterations of the \texttt{parallelproj} MLEM algorithm via the AIRPET interface. The reconstructed image is shown in Fig. \ref{fig:reco_img}. The normalization was done via division by a sensitivity matrix produced by simulating 2$\times$10$^7$ random LORs and applying an attenuation factor to account for the presence of the water ($\mu = 0.096$ cm$^{-1}$). 

\section{Future Plans}
While the current version of AIRPET provides a strong base for the functionality outlined above, it is still under development. Here we enumerate some ideas for additions in the near future:
\begin{enumerate}
    \item[-] \textbf{Additional reconstruction features:} In addition to MLEM, the present PET reconstruction techniques could be augmented with additional algorithms such as Filtered Back-Projection (FBP)~\cite{Gong2021} and the addition of time-of-flight (TOF) and scattering corrections. The construction of LORs can also be improved with the addition of detector-specific effects (TOF resolution, energy resolution and selection, etc.).
    \item[-] \textbf{Expanded AI ``tool use'':} The current AI integration is largely based on the strategy of prompting the selected LLM to write JSON corresponding to geometry elements directly. In many cases the LLM may perform better if given parameterized functions (``tools'') that it can call rather than having to invent all aspects of the geometry from primitive objects. For example, one function that is already available in AIRPET places a previously defined crystal in a ring pattern given a radius and total number of crystals, and then creates a specified number of these rings. The AI is instructed to use this function as a ``tool,'' generating results that are more reliable than if it were to attempt to correctly arrange the crystals one by one.
    \item[-] \textbf{Phantom and component libraries:} The idea would be to provide users with a repository of common PET scanner elements and standard medical phantoms, such as the Jaszczak phantom. A customizable full-body phantom could also be developed for definition of arbitrary radiotracer distributions.
    \item[-] \textbf{AI-based image evaluation:} A longer-term future goal would be to employ pre-trained medical image evaluation models such as MedGemma~\cite{Sellergren2025} or LLaVA-Med~\cite{Li2023} to interpret AIRPET-reconstructed images.
\end{enumerate}

\section*{Code availability:}
AIRPET is available as open source software (MIT license) at: \url{https://github.com/jerenner/airpet}.

\section*{Acknowledgements:} JR acknowledges support
from the Generalitat Valenciana of Spain under grant
CIDEXG/2023/16.

\bibliographystyle{JHEP}
\bibliography{refs}

\end{document}